\documentclass[aps,prd,preprint,tightenlines,superscriptaddress,showpacs,byrevtex]{revtex4}
\usepackage{graphicx}
\usepackage{dcolumn}
\usepackage{bm}

\begin{document}

\vspace*{-3\baselineskip}
\resizebox{!}{3cm}{\includegraphics{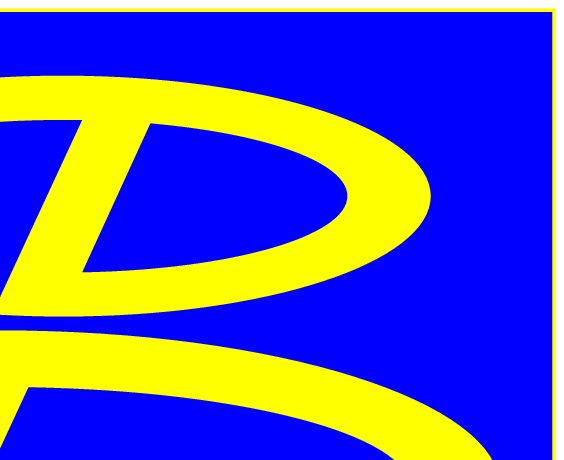}}

\preprint{\tighten\vbox{\hbox{\hfil KEK preprint 2003-110}
                        \hbox{\hfil Belle preprint 2004-1 }
}}

\newcommand{\ThisLum}{140~{\rm fb}^{-1}}
\newcommand{\LastLum}{78~{\rm fb}^{-1}}
\newcommand{\atoyerr}{\pm 0.15}
\newcommand{\stoyerr}{\pm 0.21}
\newcommand{\svalue}{-1.00}
\newcommand{\svaluecnst}{-0.84}
\newcommand{\sstaterr}{^{+0.22}_{-0.20}}
\newcommand{\ssyserr}{\pm 0.07}
\newcommand{\sresult}{\svalue~\stoyerr({\rm stat})~\ssyserr({\rm syst})}
\newcommand{\cvalue}{+0.58}
\newcommand{\cvaluecnst}{+0.55}
\newcommand{\cstaterr}{^{+0.15}_{-0.16}}
\newcommand{\csyserr}{\pm 0.07}
\newcommand{\cresult}{\cvalue~\atoyerr({\rm stat})~\csyserr({\rm syst})}
\newcommand{\avalue}{\cvalue}
\newcommand{\avaluecnst}{\cvaluecnst}
\newcommand{\astaterr}{\cstaterr}
\newcommand{\asyserr}{\csyserr}
\newcommand{\aresult}{\cresult}
\newcommand{\rt}{\rightarrow}
\newcommand{\pipi}{\pi\pi}
\newcommand{\spipi}{{\cal S}_{\pipi}}
\newcommand{\apipi}{{\cal A}_{\pipi}}
\newcommand{\rpipi}{{\cal R}}
\newcommand{\skpi}{{\cal S}_{K\pi}}
\newcommand{\akpi}{{\cal A}_{K\pi}}
\newcommand{\taub}{{\tau}_{B^0}}
\newcommand{\bz}{B^0}
\newcommand{\Bz}{\bz}
\newcommand{\bzb}{\overline{B}{}^0}
\newcommand{\dmd}{\Delta m_d}
\newcommand{\cldd}{99.999976\%}
\newcommand{\cla}{0.993}
\newcommand{\cls}{0.997}
\newcommand{\MBC}{M_{\rm bc}}

\newcommand{\apipif}{{\cal A}_{\pipi}}
\newcommand{\spipif}{{\cal S}_{\pipi}}
\newcommand{\akpif}{{\cal A}_{K\pi}}
\newcommand{\skpif}{{\cal S}_{K\pi}}
\newcommand{\abkgf}{{\cal A}_{\rm bkg}}
\newcommand{\dspipif}{\Delta \spipif}
\newcommand{\xapipi}{x_{{\cal A}\pi\pi}}
\newcommand{\xspipi}{x_{{\cal S}\pi\pi}}
\newcommand{\xamin}{x_{{\cal A}1}}
\newcommand{\xamax}{x_{{\cal A}2}}
\newcommand{\abest}{{\cal A}_{\rm best}}

\newcommand{\KLR}{LR}
\newcommand{\KLRHI}{\KLR > 0.86}
\newcommand{\KLRLO}{\KLR \leq 0.86}
\newcommand{\dE}{\Delta E}
\newcommand{\dt}{\Delta t}
\newcommand{\qq}{q\overline{q}}
\newcommand{\ks}{K_S^0}
\newcommand{\sinbb}{{\sin2\phi_1}}
\newcommand*{\fq}{\ensuremath{q}}
\newcommand{\OmCldd}{2.5\times{10^{-7}}}
\newcommand{\sigdd}{5.2\sigma}
\newcommand{\OmClda}{1.3\times{10^{-3}}}
\newcommand{\sigda}{3.2\sigma}
\newcommand{\OmClsw}{8.4\times{10^{-4}}}
\newcommand{\sigsw}{3.3\sigma}

%
%
\title{  \Large Observation of Large
     {\boldmath $CP$} Violation and Evidence for Direct
     {\boldmath $CP$} Violation 
     in $B^0 \to \pi^+\pi^-$ Decays}

\affiliation{Budker Institute of Nuclear Physics, Novosibirsk}
\affiliation{Chiba University, Chiba}
\affiliation{University of Frankfurt, Frankfurt}
\affiliation{University of Hawaii, Honolulu, Hawaii 96822}
\affiliation{High Energy Accelerator Research Organization (KEK), Tsukuba}
\affiliation{Hiroshima Institute of Technology, Hiroshima}
\affiliation{Institute of High Energy Physics, Chinese Academy of Sciences, Beijing}
\affiliation{Institute of High Energy Physics, Vienna}
\affiliation{Institute for Theoretical and Experimental Physics, Moscow}
\affiliation{J. Stefan Institute, Ljubljana}
\affiliation{Kanagawa University, Yokohama}
\affiliation{Korea University, Seoul}
\affiliation{Kyungpook National University, Taegu}
\affiliation{Swiss Federal Institute of Technology of Lausanne, EPFL, Lausanne}
\affiliation{University of Ljubljana, Ljubljana}
\affiliation{University of Maribor, Maribor}
\affiliation{University of Melbourne, Victoria}
\affiliation{Nagoya University, Nagoya}
\affiliation{Nara Women's University, Nara}
\affiliation{National United University, Miao Li}
\affiliation{Department of Physics, National Taiwan University, Taipei}
\affiliation{H. Niewodniczanski Institute of Nuclear Physics, Krakow}
\affiliation{Nihon Dental College, Niigata}
\affiliation{Niigata University, Niigata}
\affiliation{Osaka City University, Osaka}
\affiliation{Osaka University, Osaka}
\affiliation{Panjab University, Chandigarh}
\affiliation{Peking University, Beijing}
\affiliation{Princeton University, Princeton, New Jersey 08545}
\affiliation{RIKEN BNL Research Center, Upton, New York 11973}
\affiliation{Saga University, Saga}
\affiliation{University of Science and Technology of China, Hefei}
\affiliation{Seoul National University, Seoul}
\affiliation{Sungkyunkwan University, Suwon}
\affiliation{University of Sydney, Sydney NSW}
\affiliation{Tata Institute of Fundamental Research, Bombay}
\affiliation{Toho University, Funabashi}
\affiliation{Tohoku Gakuin University, Tagajo}
\affiliation{Tohoku University, Sendai}
\affiliation{Department of Physics, University of Tokyo, Tokyo}
\affiliation{Tokyo Institute of Technology, Tokyo}
\affiliation{Tokyo Metropolitan University, Tokyo}
\affiliation{Tokyo University of Agriculture and Technology, Tokyo}
\affiliation{Toyama National College of Maritime Technology, Toyama}
\affiliation{University of Tsukuba, Tsukuba}
\affiliation{Virginia Polytechnic Institute and State University, Blacksburg, Virginia 24061}
\affiliation{Yokkaichi University, Yokkaichi}
\affiliation{Yonsei University, Seoul}
  \author{K.~Abe}\affiliation{High Energy Accelerator Research Organization (KEK), Tsukuba} 
  \author{K.~Abe}\affiliation{Tohoku Gakuin University, Tagajo} 
  \author{N.~Abe}\affiliation{Tokyo Institute of Technology, Tokyo} 
  \author{T.~Abe}\affiliation{High Energy Accelerator Research Organization (KEK), Tsukuba} 
  \author{I.~Adachi}\affiliation{High Energy Accelerator Research Organization (KEK), Tsukuba} 
  \author{H.~Aihara}\affiliation{Department of Physics, University of Tokyo, Tokyo} 
  \author{K.~Akai}\affiliation{High Energy Accelerator Research Organization (KEK), Tsukuba} 
  \author{M.~Akatsu}\affiliation{Nagoya University, Nagoya} 
  \author{M.~Akemoto}\affiliation{High Energy Accelerator Research Organization (KEK), Tsukuba} 
  \author{Y.~Asano}\affiliation{University of Tsukuba, Tsukuba} 
% ...
\collaboration{The Belle Collaboration}

\date{\today}

\begin{abstract}
We report the first observation of
$CP$ violation in $B^0 \rightarrow \pi^+\pi^-$ decays
based on
152 million $\Upsilon(4S) \to B\overline{B}$ decays collected
with the Belle detector at the KEKB asymmetric-energy $e^+e^-$ collider.
We reconstruct 
a  $B^0 \rightarrow \pi^+\pi^-$
$CP$ eigenstate and identify
the flavor of the accompanying $B$ meson
from its decay products.
From the distribution of the time intervals between the two $B$ meson decay 
points, we obtain
$\apipif = \aresult$
and
$\spipif = \sresult$.
We rule out the $CP$-conserving case, $\apipi=\spipi=0$, at 
a level of 5.2 standard deviations.
We also find evidence for direct $CP$ violation with a significance 
at or greater than 3.2 standard deviations for any $\spipi$ value.

\end{abstract}


\maketitle

In the standard model (SM) of elementary particles,
$CP$ violation arises from 
the Kobayashi-Maskawa (KM) phase~\cite{KM} 
in the weak interaction quark-mixing matrix. 
In particular, the SM predicts $CP$ asymmetries 
in the time-dependent rates for $\bz$ and
$\bzb$ decays to a common $CP$ eigenstate~\cite{Sanda}. 
Comparison between SM expectations and measurements in various $CP$ eigenstates
is important to test the KM model.
The $\bz \to \pi^+\pi^-$ decay~\cite{CC},
which is dominated by the $b \to u\overline{u}d$ transition,
is of particular interest,
and is sensitive to the $CP$-violating parameter $\phi_2$.
Direct $CP$ violation may also occur in this decay 
because of interference between the $b \to u$ tree ($T$) and 
$b \to d$ penguin ($P$) amplitudes~\cite{pipipenguin}.

In the decay chain $\Upsilon(4S)\to \bz\bzb \to (\pi^+\pi^-)f_{\rm tag}$,
where one of the $B$ mesons decays at time $t_{\pi\pi}$ to the $CP$ 
eigenstate $\pi^+\pi^-$ 
and the other decays at time $t_{\rm tag}$ to a final state  
$f_{\rm tag}$ that distinguishes between $B^0$ and $\bzb$, 
the decay rate has a time dependence
given by~\cite{Sanda}
\begin{eqnarray}
\label{eq:R_q}
{\cal P}_{\pi\pi}(\Delta{t}) = 
\frac{e^{-|\Delta{t}|/{\taub}}}{4{\taub}}
\left[1 + q\cdot 
\left\{ \spipi\sin(\dmd\Delta{t})   \right. \right. \nonumber \\
\left. \left.
   + \apipi\cos(\dmd\Delta{t})
\right\}
\right],
\end{eqnarray}
where $\taub$ is the $B^0$ lifetime, $\dmd$ is the mass difference 
between the two $B^0$ mass
eigenstates, $\dt$ = $t_{\pi\pi}$ $-$ $t_{\rm tag}$, and
the $b$-flavor charge $\fq$ = +1 ($-1$) when the tagging $B$ meson
is a $B^0$ 
($\bzb$).
$\spipi$ and $\apipi$ are mixing-induced and direct $CP$-violating parameters, respectively.

Belle's previous results for $B^0 \to \pi^+\pi^-$~\cite{Acp_pipi_Belle}, 
based on a $\LastLum$ data sample (85 $\times$ 10$^6$ $B\overline{B}$ pairs),
suggested large direct $CP$ 
asymmetry and/or mixing-induced asymmetry
while the result by the BaBar collaboration
based on a sample of 88 $\times$ 10$^6$ $B\overline{B}$ pairs
did not~\cite{apipi_BaBar}.
In this Letter,
we report a new measurement
with an improved analysis that incorporates an additional 62 fb$^{-1}$ for a total of $\ThisLum$
(152$\times$10$^6$ $B\overline{B}$ pairs) that confirms
Belle's previous results with much greater significance.

The data were collected with
the Belle detector~\cite{Belle} at the KEKB asymmetric-energy
$e^+e^-$ collider~\cite{KEKB}, which collides 8.0 GeV $e^-$ 
and 3.5 GeV $e^+$ beams.
The $\Upsilon(4S)$ is produced
with a Lorentz boost of $\beta\gamma=0.425$ nearly along
the electron beamline ($z$).
Since the $B^0$ and $\bzb$ mesons are approximately at 
rest in the $\Upsilon(4S)$ center-of-mass system (cms),
$\Delta t$ can be determined from $\Delta z$,
the displacement in $z$ between the $\pi^+\pi^-$ and $f_{\rm tag}$ decay vertices:
$\Delta t \simeq (z_{\pi\pi} - z_{\rm tag})/\beta\gamma c
 \equiv \Delta z/\beta\gamma c$.
The reconstruction method of the vertex positions remains unchanged
from the previous publication~\cite{Acp_pipi_Belle}.


We use oppositely charged track pairs 
that are positively identified as pions to reconstruct
$B^0 \to \pi^+\pi^-$ candidates.
The pion efficiency is 91$\%$ and 
10.4$\%$ of kaons are misidentified as pions.
We select the $B$ meson candidates using
the energy difference 
$\Delta E\equiv E_B^{\rm cms} - E_{\rm beam}^{\rm cms}$
and the beam-energy constrained
mass $M_{\rm bc}\equiv\sqrt{(E_{\rm beam}^{\rm cms})^2-(p_B^{\rm cms})^2}$,
where $E_{\rm beam}^{\rm cms}$ is the cms beam energy,
and $E_B^{\rm cms}$ and $p_B^{\rm cms}$ are the cms energy and momentum
of the $B$ candidate.
The signal region is defined as 
$5.271 ~{\rm GeV/c^2}<\MBC<5.287 ~{\rm GeV/c^2}$
and $|\Delta{E}|<0.064$ ~GeV, corresponding to $\pm{3}\sigma$ from
the central values.
To suppress the $e^+e^- \rightarrow q\overline{q}$
continuum background ($q = u,~d,~s,~c$),  we form signal and background
likelihood functions, ${\cal L}_S$ and ${\cal L}_{BG}$, 
from the event topology, and impose requirements on the likelihood ratio
$LR$ =  ${\cal L}_S/({\cal L}_S+{\cal L}_{BG})$
for candidate events. We use the same event topology variables and the
procedure that were used for the ${\cal B}(\bz \to \pi^0\pi^0)$ 
measurement~\cite{bib:pi0pi0}.

The flavor of the accompanying $B$ meson is identified from 
inclusive properties of particles
that are not associated with the reconstructed 
$B^0 \rightarrow \pi^+\pi^-$ decay.
We use two parameters, $q$ [defined in Eq.~(\ref{eq:R_q})]
and $r$, to represent the tagging information.
The parameter $r$ is an event-by-event,
MC-determined flavor-tagging dilution factor 
that ranges from $r=0$ for no flavor
discrimination to $r=1$ for unambiguous flavor assignment.
It is used only to sort data into six $r$ intervals.
The wrong tag fractions for the six $r$ intervals, $w_l\ (l=1,6)$,
and differences between $\Bz$ and $\bzb$ decays,
$\Delta w_l$, are determined from data~\cite{bib:BC353}.


We optimize the expected sensitivity by using the improved 
likelihood ratio $LR$.
We require $\KLRHI$ for all $r$ intervals.
We include additional candidate events with  lower signal likelihood ratio
cuts (0.50, 0.45, 0.45, 0.45, 0.45, and 0.20)
for different $r$ intervals
since the separation of continuum background from the $B$ signal 
varies with $r$;
we accept candidate events from twelve distinct regions in the $LR$-$r$ plane.

\begin{figure}[!htb]
\begin{center}
\resizebox{0.6\textwidth}{!}{\includegraphics{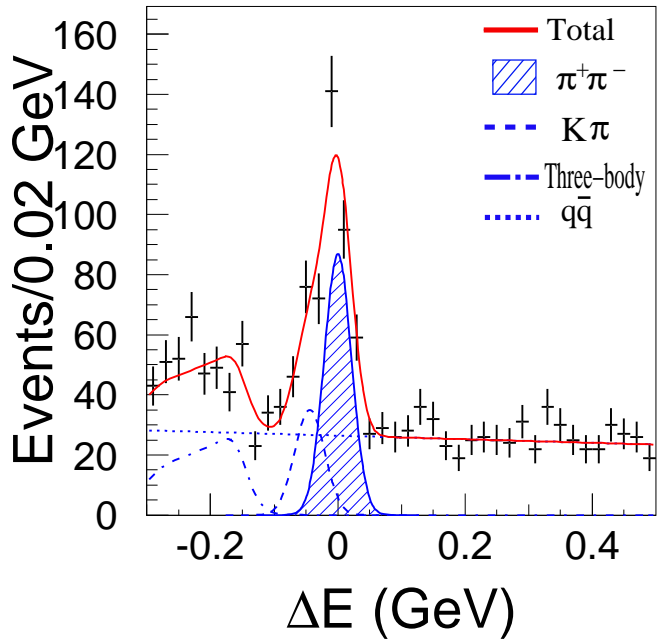}}
\end{center}
\caption{$\Delta E$ distribution in the $M_{\rm bc}$ signal region for 
$B^0 \rightarrow \pi^+\pi^-$ candidates with $\KLRHI$.}
\label{fig:DeltaE}
\end{figure}
Figure~\ref{fig:DeltaE} shows the $\Delta E$ distribution 
for the $B^0$ $\rightarrow$ $\pi^+\pi^-$ candidates 
that are in the $M_{\rm bc}$ signal region 
with $\KLRHI$ after flavor tagging and vertex reconstruction.
In the $M_{\rm bc}$ and $\Delta E$ signal region, 
we find 483 candidates with $\KLRHI$ and 1046 candidates with $\KLRLO$.
The $B^0$ $\rightarrow$ $\pi^+\pi^-$ signal yield for $\KLRHI$
is determined from an unbinned two-dimensional maximum likelihood fit to 
the $\MBC$-$\dE$ distribution
($5.20~{\rm GeV/c^2}<\MBC<5.30~{\rm GeV/c^2}$ and $-0.3~{\rm GeV}<\Delta{E}<0.5$ ~GeV )
with a Gaussian signal function
plus contributions from misidentified $B^0 \to K^+\pi^-$ events, 
three-body $B$-decays, and continuum background. 
The fit yields 
$232^{+20}_{-19}$ $\pi^+\pi^-$ events and $82^{+14}_{-13}$ $K^+\pi^-$ 
events in the signal region, where the errors are statistical only.
Extrapolating from the size of the continuum background in this fit,
we expect $169$ continuum events in the signal region.
For $\KLRLO$, 
the same procedure used in the previous publication~\cite{Acp_pipi_Belle} yields
$141\pm 12$ $\pi^+\pi^-$ events, $50\pm 8$ $K^+\pi^-$ events and
$855$ continuum events in the signal region.
The contribution from
three-body $B$-decays is negligibly small in the signal region.

The $\dt$ resolution function $R_{\pi\pi}$ for
$B^0$ $\rightarrow$ $\pi^+\pi^-$ signal events is formed by convolving four components:
the detector resolutions for $z_{\pi\pi}$ and $z_{\rm tag}$,
the shift in the $z_{\rm tag}$ vertex position due to secondary tracks originating
from charmed particle decays,
and the smearing due to the kinematic approximation used to convert $\Delta z$
to $\Delta t$~\cite{bib:BC353}.
We assume $R_{\pi\pi}$ = $R_{K\pi}$ and denote them
collectively as $R_{\rm sig}$.

$\apipif$ and $\spipif$
are obtained from an unbinned maximum likelihood fit
to the observed $\dt$ distribution. 
The probability density function (PDF) for $B^0$ $\rightarrow$ $\pi^+\pi^-$ signal events (${\cal P}^q_{\pi\pi}$) 
is given by
Eq.(\ref{eq:R_q}) modified to incorporate the effect of incorrect flavor assignment.
The PDF for 
$B^0$ $\rightarrow$ $K^+\pi^-$ background events is
${\cal P}^q_{K\pi}(\Delta t,w_l, {\Delta w_l})
  =
  \frac{1}{4\tau_{B^0}}{ e^{-|\Delta t|/\tau_{B^0}} }
 \{
 1 - q{\Delta w_l}+ q\cdot (1-2w_l)\cdot{\akpif}
 \cdot\cos(\Delta m_d\Delta t)
 \}.$
We use $\akpif = 0$ as a default and include an effect of
possible non-zero value for $\akpif$ in the systematic error.
The PDF for continuum background events is 
${\cal P}_{q\overline{q}}(\Delta t)
 =
(1+q{\cdot}{\abkgf})\{\frac{f_\tau}{2\tau_{\rm bkg}}e^{-|\Delta t|/\tau_{\rm bkg}}
 + ( 1 - f_\tau )\delta(\Delta t)\}/2,$
where $f_\tau$ is the fraction of the background with effective lifetime
$\tau_{\rm bkg}$ and $\delta$ is the Dirac delta function. 
We use $\abkgf = 0$ as a default.
A fit to sideband events yields
$\abkgf = 0.010 \pm 0.005$. This uncertainty is included in
the systematic error for $\apipi$ and $\spipi$.
All parameters of ${\cal P}_{\qq}(\Delta t)$ and $R_{\qq}$ are
determined from the events in the
sideband region.

We define the likelihood value for each ($i$th) event as 
a function of $\apipif$ and $\spipif$:
\begin{eqnarray}
{P_i =
(1 - f_{ol})\int^{+\infty}_{-\infty}
[\{f^m_{\pi\pi}{\cal P}^q_{\pi\pi}(\Delta t^\prime, w_l; \apipif, \spipif)} \nonumber \\
+ f^m_{K\pi}{\cal P}^q_{K\pi}(\Delta t^\prime, w_l)\}
\cdot R_{\rm sig}(\Delta t_i-\Delta t^\prime) \nonumber \\
  + f^m_{q\overline{q}}{\cal P}_{q\overline{q}}(\Delta t^\prime)
\cdot R_{q\overline{q}}(\Delta t_i-\Delta t^\prime)]
d\Delta t^\prime 
+ f_{ol}{\cal P}_{ol}(\Delta{t_i}).
\label{eq:likelihood}
\end{eqnarray}
Here the probability functions $f^m_k$ ($k$ = $\pi\pi$, $K\pi$ or 
$q\overline{q}$ )
are determined on an event-by-event basis as functions of  $\Delta E$ 
and $\MBC$ for each $LR$-$r$ interval ($m$= 1, 12)~\cite{Acp_pipi_Belle}. 
The small number of signal and background 
events that have large values of $\Delta{t}$ 
are accommodated by the
outlier PDF, ${\cal P}_{ol}$, with
fractional area $f_{ol}$.
In the fit,
$\spipif$ and $\apipif$ are the only free parameters 
determined by maximizing
the likelihood function
${\cal L}=\prod_i P_i$, where the product is over all
$B^0 \rightarrow \pi^+\pi^-$ candidates.

\begin{figure}[!htbp]
\begin{center}
\resizebox{0.6\textwidth}{!}{\includegraphics{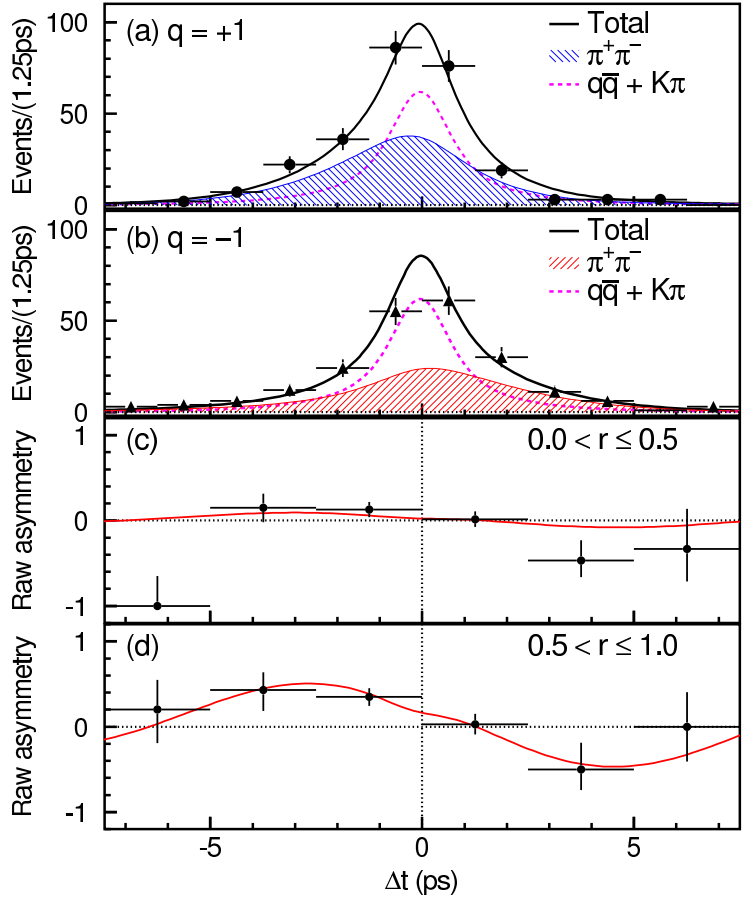}}
\end{center}
\caption{
The $\Delta t$ distributions for 
the 483 $B^0 \rightarrow \pi^+\pi^-$ candidates with $\KLRHI$ in the
signal region: 
(a) 264 candidates with $q = +1$, i.e. the tag side is identified as $B^0$;
(b) 219 candidates with $q = -1$.
(c) Asymmetry, $A$, in each $\dt$ bin with $0 < r \leq 0.5$ and 
(d) with $0.5 < r \leq 1.0$.
The solid curves show the
results of the unbinned maximum likelihood fit to the $\Delta t$
distributions of the 1529 $B^0$ $\rightarrow$ $\pi^+\pi^-$ candidates.
}
\label{fig:asym}
\end{figure}

The unbinned maximum likelihood fit to the
1529 $B^0$ $\rightarrow$ $\pi^+\pi^-$ candidates 
(801 $B^0$- and 728 $\bzb$-tags), 
containing 372$^{+32}_{-31}$ $\pi^+\pi^-$ signal events, yields
$\apipif = \aresult$ and $\spipif = \sresult$.
The correlation between $\apipif$ and $\spipif$ is 0.286.
As in our previous publication~\cite{Acp_pipi_Belle},
we quote the rms values of the $\apipif$ and $\spipif$
distributions of the MC pseudo-experiments
as the statistical errors of our measurement~\cite{footnoteErrorReduction}. 
The usual fit errors from the likelihood functions, 
called the MINOS errors in the previous
publication~\cite{Acp_pipi_Belle}, are $\cstaterr$ and $\sstaterr$ for
$\apipi$ and $\spipi$, respectively, in good agreement with the rms 
values above~\cite{footnoteStatError}.
In Figs.~\ref{fig:asym}(a) and (b), we show the 
$\Delta t$ distributions for
the 264 $B^0$- and 219 $\bzb$-tagged events in the subset of data with $\KLRHI$.
We define the raw asymmetry in each $\dt$ bin by 
$A \equiv (N_{+} - N_{-})/(N_{+} + N_{-})$, where 
$N_{+(-)}$ is the number of observed candidates with $q = +1(-1)$.
Figures~\ref{fig:asym}(c) and (d) show the raw asymmetries for two regions of 
the flavor-tagging parameter $r$.
The effective tagging efficiency and signal purity is much larger 
in the $0.5 < r \leq 1.0$ region.

We test the goodness-of-fit from a $\chi^2$
comparison of the results of the unbinned fit and
the $\Delta t$ projections for $B^0 \rightarrow \pi^+\pi^-$ candidates.
We obtain $\chi^2$/DOF = 12.5/12 (7.6/12)
for the $\Delta t$ distribution of
the $B^0$ ($\bzb$) tags.

An ensemble of MC pseudo-experiments indicates a 26.7$\%$ probability 
of measuring $CP$ violation at a level above the one we observe when the input values
are $\apipi = \avaluecnst$ and $\spipi = \svaluecnst$, 
which correspond to the values at the point of maximum likelihood 
in the physically allowed region ($\spipi^2+\apipi^2\leq 1$); in this measurement it is
located at the physical boundary ($\apipi^2 + \spipi^2 = 1$).

The systematic error is primarily due to
uncertainties in the vertexing ($\pm 0.04$ for $\apipif$ and $\pm 0.05$ for $\spipif$) and 
the background fractions ($\pm 0.03$ for $\apipif$ and $\pm 0.02$ for $\spipif$). 
We include the effect of tag side interference~\cite{bib:TSI} on
$\apipi (\pm 0.03)$ and $\spipi (\pm 0.01)$ in this analysis.
Other sources of systematic error are uncertainties in the wrong tag fraction, physics parameters
($\Delta m_d$, $\tau_{B^0}$, and $\akpi$), resolution function, background modeling, and fit bias.
We add each contribution in quadrature to obtain the total systematic errors.
The effect of the 3\% charge asymmetry in the kaon misidentification rate
is negligibly small.

We perform a number of crosschecks. 
We measure the $B^0$ lifetime with
the $B^0 \rightarrow \pi^+\pi^-$ candidate events.
The result, 
$\tau_{B^0} = 1.46\pm 0.09{\rm ~ps}$,
is consistent with the world-average value~\cite{PDG}.
A comparison of the event yields and $\Delta t$ distributions for
$B^0$- and $\bzb$-tagged events in the sideband region reveals
no significant asymmetry.
We select $B^0 \to K^+\pi^-$ candidates
by positively identifying the charged kaons.
A fit to the 2358 
candidates (1198 signal events)
yields $\akpif = -0.02 \pm 0.08$, in agreement with the
counting analysis~\cite{dcpv_lp03},
and $\skpif = 0.14 \pm 0.11$, which is consistent with zero.
With the $K^+\pi^-$ event sample, 
we determine
$\taub = 1.52\pm 0.06$~ps 
and 
$\dmd = 0.53^{+0.04}_{-0.07}$~ps$^{-1}$, which are
in agreement with the world average values~\cite{PDG}.
We check the measurement of $\apipif$ using time-independent fits to 
the $\MBC$-$\Delta E$ distributions for the $B^0$ and $\bzb$ tags.
We obtain $\apipif = +0.73\pm 0.19$, which is consistent with 
the time-dependent $CP$ fit result.
We also perform an independent analysis based on a binned maximum-likelihood fit to 
the $\dt$ distribution.
The result is consistent with that of the unbinned maximum-likelihood fit quoted here.

\begin{figure}[!htb]
\resizebox{0.6\textwidth}{!}{\includegraphics{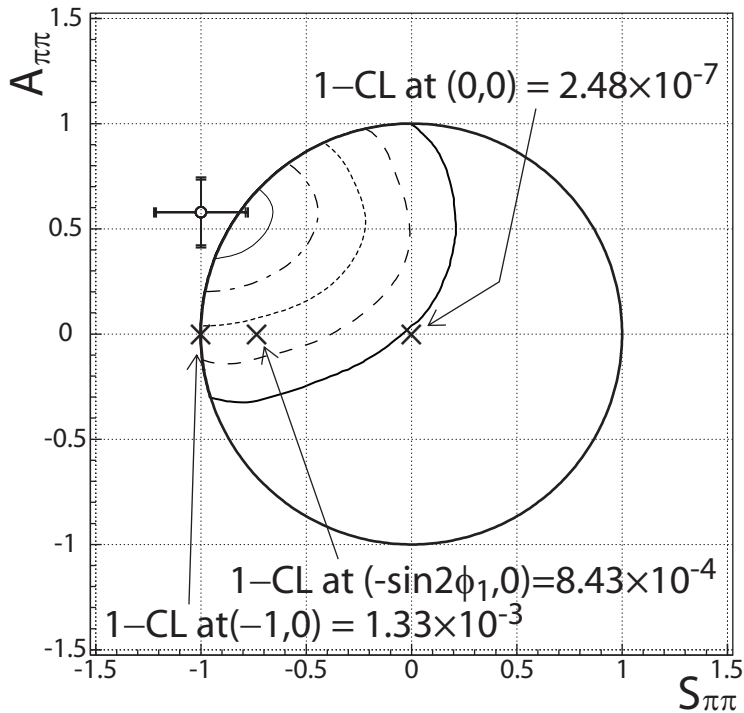}}
\caption{Confidence regions for $\apipi$ and $\spipi$.
The curves show the contours for 1-CL = $3.17 \times 10^{-1}$ (solid),
$4.55 \times 10^{-2}$ (dot-dashed), $2.70 \times 10^{-3}$ (dotted),
$6.34 \times 10^{-5}$ (dashed), and $5.96 \times 10^{-7}$ (thick solid).}
\label{fig:2dcl}
\end{figure}

The statistical significance of our measurement is determined from
the same approach used in the previous publication~\cite{Acp_pipi_Belle}.
Figure~\ref{fig:2dcl} shows the resulting two-dimensional confidence regions 
in the $\apipi$ vs. $\spipi$ plane.
The case that $CP$ symmetry is conserved, $\apipi=\spipi=0$, is ruled out
at the $\cldd$ confidence level (CL), i.e., $1-{\rm CL} = \OmCldd$, 
equivalent to $\sigdd$ significance for Gaussian errors.
The case of no direct $CP$ violation, $\apipi$ = 0, is also ruled out
with a significance at or greater than 
$\sigda$ for any $\spipi$ value.
%
If the source of $CP$ violation is only due to $B$-$\overline{B}$ mixing
or $\Delta B=2$ transitions as in so-called 
superweak scenarios~\cite{bib:superweak},
then ($\spipi, \apipi$) = ($-{\rm sin}2\phi_1, 0$).
$1-{\rm CL}$ at this point is $\OmClsw$, equivalent to $\sigsw$ significance.

Adopting the notation of Ref.~\cite{bib:SA_TH_GR}, 
the range of $\phi_2$ that corresponds to the 95.5$\%$ CL region for
$\apipi$ and $\spipi$ in Fig.~\ref{fig:2dcl} 
is $90^\circ \leq \phi_2 \leq 146^\circ$
for $0.15<|P/T|<0.45$ as used in the previous publication~\cite{Acp_pipi_Belle} and 
$\sin{2\phi_1}=0.736$~\cite{bib:phi1}. 
The result is in agreement with constraints on the unitarity triangle
from other indirect measurements~\cite{bib:Schubert}.
The 95.5\% CL region for $\apipi$ and $\spipi$ excludes $|P/T|<0.17$.

In summary, we have performed a new measurement of
$CP$ violation parameters in $B^0 \rightarrow \pi^+\pi^-$ decays.
We obtain $\apipif = \aresult$, and $\spipif = \sresult$.
We rule out the $CP$-conserving case, $\apipi=\spipi=0$, at 
the $\sigdd$ level.
We find evidence for direct $CP$ violation with a significance
at or greater than $\sigda$.
The constraints on $\phi_2$ from our result are consistent with
indirect measurements that assume the correctness of the SM.


  We thank the KEKB group for the excellent
  operation of the accelerator, the KEK Cryogenics
  group for the efficient operation of the solenoid,
  and the KEK computer group and the NII for valuable computing and
  Super-SINET network support.  We acknowledge support from
  MEXT and JSPS (Japan); ARC and DEST (Australia); NSFC (contract
  No.~10175071, China); DST (India); the BK21 program of MOEHRD and the
  CHEP SRC program of KOSEF (Korea); KBN (contract No.~2P03B 01324,
  Poland); MIST (Russia); MESS (Slovenia); NSC and MOE (Taiwan); and DOE
  (USA).


\begin{thebibliography}{99}
\bibitem{CC}
Throughout this paper, 
the inclusion of the charge conjugate mode decay is implied
unless otherwise stated.
\bibitem{footnoteErrorReduction}
The statistical errors are reduced by about a factor of two due to 
the increase in the data sample (0.75), the improvement in signal yield extraction (0.90)
and continuum rejection (0.95), the use of a more accurate signal fraction (0.93) and
other effects (0.93). The values in parentheses are the expected reduction factors.
\bibitem{footnoteStatError}
The log-likelihood ratio $-2{\rm ln}{\cal L}/{\cal L}_{max}$ 
is now nearly parabolic and 
the sensitivity  to a large single event fluctuation~\cite{Acp_pipi_Belle} is 
greatly reduced.
\end{thebibliography}
\end{document}